\documentstyle[aps,epsfig,amssymb,preprint,tighten]{revtex}

\newcommand{\be}{\begin{equation}}
\newcommand{\ee}{\end{equation}}
\newcommand{\bea}{\begin{eqnarray}}
\newcommand{\eea}{\end{eqnarray}}

\begin{document}
\title{Glueball masses in 4d U(1) lattice gauge theory using the multi-level 
algorithm}
\author{Pushan Majumdar\thanks{e-mail:pushan@mppmu.mpg.de}\and
Yoshiaki Koma \thanks{e-mail:ykoma@mppmu.mpg.de}
\and Miho Koma \thanks{e-mail:mkoma@mppmu.mpg.de}}
\address{Max-Planck-Institut f\"ur Physik, F\"ohringer Ring 6,
D-80805, M\"unchen, Germany.}
\preprint{Preprint No. : MPP-2003-69}
\maketitle
\begin{abstract}
We take a new look at plaquette-plaquette correlators in 4d compact U(1)
lattice gauge theory which are separated in time, both in the confined 
and the deconfined phases. From the behaviour of these correlators we 
 extract glueball masses in the scalar as well as the axial-vector channels.
Also in the deconfined phase, the non-zero momentum axial-vector correlator
gives us information about the photon which appears as a massless particle in 
the spectrum. 
Using the L\"uscher - Weisz multi-level algorithm, we are able to go to large
 time separations which were not possible previously.  
\end{abstract}

\section{Introduction}

Compact U(1) lattice gauge theory in 4d exists in two phases
separated by what is believed to be a weak first order transition. This theory 
does not have some complications of
the non-Abelian gauge groups but still exhibits confinement of test charges in 
one of the phases due to the compact nature of its dynamical
variables. Thus it is an ideal ground for testing various ideas relevant
for confinement. 

Abelian lattice gauge theory was first formulated and analyzed by Wilson in 1974
\cite{Wilson}.
He concluded that at strong coupling the theory indeed confined static
test quarks by looking at path ordered exponentials
around closed paths (Wilson loops). A plausible mechanism of confinement was
first given by Polyakov \cite{Polyakov} who showed
that in 3d confinement in Abelian lattice gauge theory could be thought
of as due to the presence of a monopole plasma which produced an area law for the
Wilson loop correlation function for all values of the coupling constant. The existence of 
monopoles even in a U(1) theory
were a direct consequence of the compact nature of the dynamical variables.
In 4d both Wilson and Polyakov conjectured that there had to be a
weakly coupled regime where test charges were not confined and thus the theory
should have at least two distinct phases. The existence of a phase transition was
finally proved by Guth \cite{Guth}.
 
Most of the analytical work was done in the ``Villain approximation" where the
original action is replaced by a quadratic action, but the periodicity of the
variables is retained by using the Poisson summation formula. It was in this
approximation that the existence of a massless photon was established in a
weak coupling regime \cite{FS}. The Villain approximation achieves a separation
between the perturbative and non-perturbative degrees of freedom. It is widely
believed that the original action and the Villain action fall in the same
universality class and therefore they should have the same critical behaviour.
Except for the strong coupling limit, the model with its original action has
mostly been studied numerically starting from \cite{Creutz}.
 
Numerically practicable proposals to detect monopoles were first discussed by
DeGrand and Toussiant \cite{DT} and now there is evidence that at strong
coupling there is a non-zero
monopole density (condensate) and beyond a certain coupling $g_c$ the monopole
density abruptly drops almost to zero. Therefore our present understanding is
that compact U(1) lattice gauge theory exists in a confining and a deconfining
phase which are separated by a phase transition.
The confinement mechanism is thought to be due to the presence
of monopoles and there are analogies to a dual superconductor mechanism \cite{super}. 
In the deconfined phase, away from
the transition region, one expects a massless photon. However nothing rigorous 
is known around the phase transition point.
 
Monte Carlo simulations have established that for the Wilson action the
transition point corresponds to $\beta=1.011128 (11)$ \cite{ALSN}. The order of the 
transition
has long been a matter of debate. Recent high statistics investigations 
including finite size scaling analysis have suggested a weak first order transition 
\cite{ALSN}. Other actions have also been studied and for the extended action which
has two couplings, there have been interesting claims of the existence of a 
second order phase transition at particular values of these couplings \cite{EA}.

Accurate measurements of the glueball mass can throw light on the order of the
phase transition. However 
correlators in compact U(1) theory are difficult to handle numerically in 
the confining phase as the 
signal to noise ratio rapidly becomes worse with increasing distance. 
In fact except for points close to the phase transition \cite{Stack}
(where the glueball
is lighter), such measurements have been carried out only for  small temporal 
extents of the correlators \cite{Oldphoton}. 

Recently L\"uscher and Weisz have proposed an exponential noise reduction method
which exploits the local nature of the action and the existence of a positive
definite transfer matrix \cite{LW1}. Using this method, exponentially small values of
 Polyakov loop correlators at large separations have been measured reliably for both SU(2) 
\cite{PM} and SU(3) \cite{LW2} gauge groups. Large Wilson loops have also been measured for 
SU(2) in 3d \cite{PM} . This gives the spectrum of the hadronic string. The breaking 
of the adjoint string \cite{F1}, the 3-quark potential \cite{F2} and the glueball spectrum 
for SU(3) gauge group \cite{Meyer} have also been studied using this procedure.
In this work we apply this error reduction procedure to 4d compact U(1) lattice gauge 
theory. We measure the scalar and axial vector glueball masses and also explore
the masslessness of the photon close to the transition point in the deconfined phase. 

The multi-level algorithm does not stipulate any fixed rule as how to measure a given
observable and has to be applied differently in a way appropriate to the observable
in question. The methods we employ for the measurements are quite new and 
we believe are of as much importance as the results
themselves. These methods let us go to large physical separations for
the correlators in question. This is very important as at large separations, 
the contamination from higher excited states are small and the signals are
relatively clean.

In section II we give the parameters of our simulation. Section III deals with the 
glueball correlators and there we explain how to use the multi-level algorithm to
measure both scalar and the axial-vector glueball correlators. In section IV we present
our results for the masses and in section V we present a discussion of our results. 

\section{Simulation details}

In our simulations we use the Wilson action given by
\be
S[U]=\beta\sum_{n,\mu,\nu}\left\{ 1 - {\bf Re}(e^{i\theta_{\mu\nu}(n)})\right\},
\ee
where $\theta_{\mu\nu}(n)$ is constructed from the dynamical variables $\theta_{\mu}(n)$
as
\be
\theta_{\mu\nu}(n)=\theta_{\mu}(n)+\theta_{\nu}(n+{\hat \mu})-\theta_{\mu}(n+{\hat 
\nu})-\theta_{\nu}(n).
\ee
In this formulation $\theta_{\mu}(n)$ 
is compact and ranges $[-\pi,\pi]$. We impose periodic boundary 
conditions in all directions. The relation with continuum 
perturbation theory is obtained by identifying $\beta$ with $1/g_0^2$, where $g_0$ is the 
bare coupling of the perturbation theory. All our simulations are done on a 
$16^4$ lattice and on this lattice the phase transition occurs around $\beta=1.0108$
\cite{ALSN}.
We do not probe the transition too closely. The values of the coupling that we look at
are given by $\beta=0.990, 1.000, 1.005, 1.010 $ in 
the confining region and $\beta=1.012, 1.015$ and $1.020$ in the deconfining region. 
To observe the photon we also simulate at $\beta = 1.03, 1.04, 1.05$ and $1.06$.
We invoke the usual heatbath algorithm \cite{Prog} and a ratio of one heatbath to three 
over-relaxations. We start from an ordered configuration (cold start) and use the 
first 1000 updates for thermalization.

To get an idea of the lattice spacing in the confining region, we use the string tension obtained 
from 
the Polyakov loop correlators \cite{Scale}. Appealing to the universality of the string picture, 
we assume that in this case $\sqrt\sigma$ also
corresponds to $(0.5 f\!m)^{-1}$. The string tension and the lattice spacings are then given 
by 

\begin{center}
\begin{tabular}{c|c|c|c|c}
$\beta$ & 0.99 & 1.0 & 1.005 & 1.01 \\
\hline
$ a^2\sigma $ & 0.231 (2) & 0.164 (2)& 0.123 (2)& 0.063 (2)\\
\hline
$ a(f\!m) $ & 0.240 (1)& 0.202 (2)& 0.175 (2)& 0.125 (2)
\end{tabular}
\end{center}

In the deconfined phase the string tension vanishes and we know that
in the weak coupling limit the theory must go over to free electrodynamics which
is a scale invariant theory. Therefore we do not attempt to set the scale in the 
deconfined regime.  

\section{Glueball correlators}

Glueballs in 4d compact U(1) lattice gauge theory were first investigated by Berg and 
Panagiotakopoulos \cite{Oldphoton}. They found that the masses in the scalar and the 
axial-vector 
channels with zero momentum fell quite sharply as one went towards the transition region
from the confining phase but started rising again as one went towards the weak 
coupling region in the deconfined phase. In stark contrast masses from the momentum 
dependent axial-vector correlator dropped dramatically in the deconfined phase. Assuming
the nearest neighbour lattice free field dispersion relation, they concluded that 
their data for the axial-vector correlator was consistent with the presence of a 
massless photon in this phase. These studies were carried out on 
a $4^3\times 8$ lattice and consequently their 
lowest momentum was still quite high at $2\pi/4$.
Glueball masses in U(1) were also looked at by Stack and Filipczyk \cite{Stack}. 
They concluded 
that at least in the scalar channel, the glueball mass could be obtained by only 
looking at the monopole part of the Wilson loop.  
Glueball masses have also been studied extensively for the extended action 
\cite{EA}. 
However for that action the transition is expected to be of second order
and therefore the critical behaviour is most likely quite different from the 
critical behaviour of the usual Wilson action.
 
\subsection{Scalar channel}
In this sub-section we explain how to measure glueball masses in the scalar channel
using the multi-level scheme. To obtain the mass of this glueball, we measure
the connected part of the correlator $\langle C(t)C(t_0) \rangle $ given by 
\be\label{corr}
\langle C(t)C(t_0) \rangle_{conn}
=\langle \sum_{{\bf n},i,j}P_{ij}({\bf n},t)\sum_{{\bf n},i,j}P_{ij}({\bf n},t_0) 
\rangle -\langle \sum_{{\bf n},i,j}P_{ij}({\bf n},t)\rangle \langle 
\sum_{{\bf n},i,j}P_{ij}({\bf n},t_0) \rangle 
\ee
where $P_{ij}({\bf n},t)=\exp (i\,\theta_{ij}({\bf n},t))$ is the plaquette in the $ij$ 
plane. ${\bf n}$ goes over all the points in a timeslice and $i,j$ goes over the values
$1,2$ and $3$.
This correlator is expected to behave like 
\begin{equation}
\langle C(t)C(t_0) \rangle_{conn}\approx \alpha \left [ 
e^{-m(t-t_0)}+e^{-m(N_t-(t-t_0))} \right ],
\end{equation}
where $N_t$ is the extent of the lattice in the time direction.
Fitting the measured correlator to this form, one can read off the mass $m$ of the 
glueball. However this naive approach is not particularly suited for the multi-level 
algorithm.

One of the most efficient uses for the multi-level algorithm is to generate the small
expectation values by multiplication rather than fine cancellation of positive
and negative values of the same order.
In equation (\ref{corr}) this advantage is lost since each expectation value is 
a number ${\cal O}(1)$, but the connected part is several orders of magnitude smaller 
than the full correlator.
To get around this problem, we can take the derivative of the correlator to get rid of 
the vacuum expectation values of the plaquettes. So let us now take the derivative of 
the correlator at both $t$ and $t_0$ \footnote{In principle one derivative is enough,
 but in practice we found that the efficiency of the algorithm is higher for the double
derivative compared to the single one.} to obtain 
\begin{equation}\label{dcorr}
\partial_t\partial_{t_0}^* \langle C(t)C(t_0) \rangle
\approx\alpha \left [\partial_t e^{-mt}\partial_{t_0}^*e^{mt_0}
+e^{-mN_t}\partial_te^{mt}\partial_{t_0}^*e^{-mt_0} \right ].
\end{equation}
Taking $\partial_t$ to be the forward derivative and $\partial_{t_0}^*$ to be the 
backward derivative on the lattice, we get,
\be\label{dcorr1}
\partial_t\partial_{t_0}^* \langle C(t)C(t_0) \rangle \approx
-\alpha \left [ e^{-m(t-t_0)}(1-e^{-m})^2 + e^{-m(N_t-(t-t_0))}(e^m-1)^2\right ].
\ee

Now we are far better suited to apply the multi-level scheme. We can now measure 
\be\label{dcorr2}
 \partial_t\partial_{t_0}^* \langle C(t)C(t_0) \rangle = \langle \sum_{{\bf n},i,j}
\left [ P_{ij}({\bf n},t+1)-P_{ij}({\bf n},t) \right ]
\sum_{{\bf n},i,j} \left [ P_{ij}({\bf n},t_0)-P_{ij}({\bf n},t_0-1) \right ] \rangle ,
\ee
where [$\cdots$] denotes the sub-lattice average of a quantity. 

Sub-lattice averages,
which were introduced in \cite{LW1}, are averages of quantities in their local
environments. This requires partial updates of the lattice. In our case, 
for example, while estimating the expectation value of the plaquettes at time $t$ (see
fig. 1), the links in the spatial directions at time slices $(t+1)$ and $(t-1)$ are held 
fixed 
while the links between these fixed boundaries are updated using the usual mixture 
of heat-bath and over-relaxation. The sub-lattice averages are obtained by averaging the 
operator over these updated links.  After several sub-lattice updates, the full lattice 
is updated as usual. The average over the sub-lattice updates constitutes one measurement
in this case. Thus a measurement in the multi-level scheme is expensive, but the measured 
values are already quite stable and show very little fluctuation. This procedure  
uses crucially the locality of the action, as all the links which are affected by an
updated link have to fit inside the slice of the lattice whose boundary is being held 
fixed during the update.
The derivatives were estimated in the sub-lattice updates by taking the difference
of the value of the operator on the updated slice with the value of the operator  
on the boundary. As shown in fig. 1, to get the forward derivative at $t$, we used the
fixed boundary at $(t+1)$ and for the backward derivative, the boundary at $(t-1)$.   
To get the correlators one has to use two such slices (e.g. $t$ and $t_0$ in
fig. 1.).
In practice we hold every alternate layer of spatial links fixed and estimate
the correlators for various time separations at the same time.
The only drawback at the moment seems to be the fact that we  
have to consider a minimum separation of two in the temporal direction.

The number of sub-lattice updates is an optimization
parameter of the algorithm that has to be tuned for efficient performance.
This is a function of $\beta$. In the range of $\beta$ we looked at, we found
that 10 to 50 sub-lattice updates were sufficient.
To compare this procedure with the naive algorithm we measured the percentage
error on the correlators at a value of $(t-t_0)$ where both methods gave non-zero
signals. In a similar amount of computer time, the multi-level algorithm produced 
errors which were about two orders of magnitude lower than the naive method.

\begin{figure}[htb]
\begin{center}
\mbox{\epsfig{file=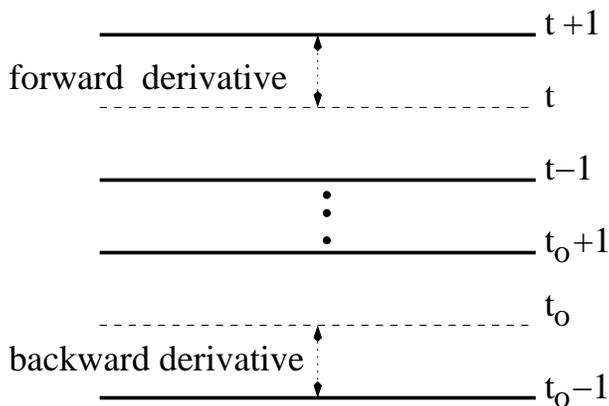,width=8truecm,angle=0}}
\caption{ Evaluation of the derivative of the glueball correlator. The thick lines
are held fixed during the sub-lattice averaging.}
\end{center}
\end{figure}

\subsection{Axial-vector channel}

For glueball masses in the axial-vector channel, we follow the same 
procedure as in the scalar channel, except instead of the full plaquette,
we now look at the correlation between the imaginary part of the space-like 
plaquettes separated by time $\Delta t$. In this channel, a zero momentum glueball
is created by
\be
B_{ij}(t)= \sum_{\bf n} {\bf Im}\:P_{ij}({\bf n},t),
\ee
where again $P_{ij}({\bf n},t)$ is the plaquette in the $ij$ plane,  
$i$ and $j$ go over ${1,2,3}$ and ${\bf n}$ goes over all the points in a time slice. 
This correlator does not have a vacuum 
expectation value. Therefore for this channel we do not have to take the derivatives
before applying the multi-level scheme. We directly estimate 
\be\label{vector}
\langle [ B_{ij}(t)][B_{ij}(t_0)]\rangle \approx \alpha 
[e^{-m_{\rm eff}(t-t_0)}+e^{-m_{\rm eff}(N_t-(t-t_0))}]
\ee 
both in the confining as well as the deconfining phase.

The imaginary part of the plaquette is like $\sin (\theta_{ij})$ and under interchange 
of the indices $ij$, it picks up a negative sign. Thus $\sin (\theta_{ij})$ can be 
considered a 2-form. In the transfer matrix formulation, the correlator of two 2-forms 
with momenta ${\bf k}$ is given by
\be\label{vcorr1}
\langle B_{ij}({\bf k},t)B_{ij}(-{\bf k},t_0)\rangle = Z^{-1} \:Tr
\left ( \sum_{\bf n} {\bf Im}\:e^{i{\bf k}\cdot{\bf n}}{\hat P}_{ij}({\bf n},t)\right ) 
{\hat T}^{(t-t_0)}
\left ( \sum_{\bf n} {\bf Im}\:e^{-i{\bf k}\cdot{\bf n}}{\hat P}_{ij}({\bf n},t_0)\right ) 
\ee
where ${\hat T}$ is the transfer matrix.
Each component of ${\bf k}$ takes the values $2\pi l/N_s$ where $N_s$ is the spatial extent 
of the lattice 
and $l$ is an integer going from $1$ to $N_s$. If $d$ is the exterior derivative, then, as in the 
continuum, $d^2=0$ also holds on the lattice \cite{dform} and 
we can rewrite $B$ as $dA + H$ where $A$ is a 1-form and $H$ is the part of $B$ that cannot
be written as $d$ of a 1-form. Inserting this decomposed version into Eq.(\ref{vcorr1}) 
we get
\be\label{vcorr3}
\langle B_{ij}B_{ij}\rangle\approx Tr \left ( \sum_{\bf n} e^{i{\bf k}\cdot{\bf 
n}}(\partial_{[i}A_{j]} + H_{ij}) \right ) e^{-m_{\rm eff}(t-t_0)}\left ( \sum_{\bf n} e^{-i{\bf 
k}\cdot{\bf n}}(\partial_{[i}A_{j]} + H_{ij}) \right ),
\ee
where we have assumed that states of a definite momentum, which are eigenstates of $B$ by
construction also diagonalise the Hamiltonian ${\cal H}$. This is certainly true in the 
weak coupling region, but we believe, is not a bad approximation to make throughout the 
deconfined 
phase. Next doing an integration by parts and recalling that we have periodic boundary 
conditions, we get
\be\label{vcorr4}
\langle B_{ij}B_{ij}\rangle\approx Tr\left ( \sum_{\bf n}A_jk_i+\sum_{\bf n} e^{i{\bf 
k}\cdot{\bf n}}H_{ij}\right ) 
\left ( \sum_{\bf n}A_jk_i+\sum_{\bf n} e^{-i{\bf k}\cdot{\bf n}}H_{ij}\right 
)e^{-m_{\rm eff}(t-t_0)}.
\ee
Thus the correlator has a momentum dependent part which is sensitive to a correlator
between two 1-forms. In the deconfining region we expect this part to yield information
about the photon which we know exists in the weak coupling regime.  It is interesting to 
observe that at ${\bf k}=0$ there is no photon contribution to the $\langle BB\rangle $
correlator Eq.(\ref{vcorr1}), hence one has to consider a non-zero total momentum to see 
the photon.

\section{Results}

\subsection{Scalar channel}

In the scalar channel we measured both $\partial_t\partial_{t_0}^* \langle
C(t)C(t_0) \rangle$ and $\partial_t^*\partial_{t_0} \langle C(t)C(t_0) \rangle$. 
The only difference between the two quantities is the interchange of coefficients
$(1-e^{-m})^2$ and $(e^m-1)^2$.
As can be easily seen from the explicit form given in Eq. (\ref{dcorr1}), 
both the functions can be rewritten 
as $\Gamma^{'}\!\cosh (m\Delta t)$. We folded the data about the symmetry point and 
performed correlated fits to this form over 
different ranges of $\Delta t$. The effective masses and $\chi^2$ from the 
fits are given in table I. 

The effective masses obtained from the two sets are consistent with each other 
within error bars. We also see that at the smallest value of $\Delta t$, the effective 
masses have significant contamination from the higher states. This effect 
becomes more and more pronounced as one approaches the phase transition.    
In fig. 2 we plot the correlation functions themselves.

\begin{figure}[]
\begin{center}
\mbox{\epsfig{file=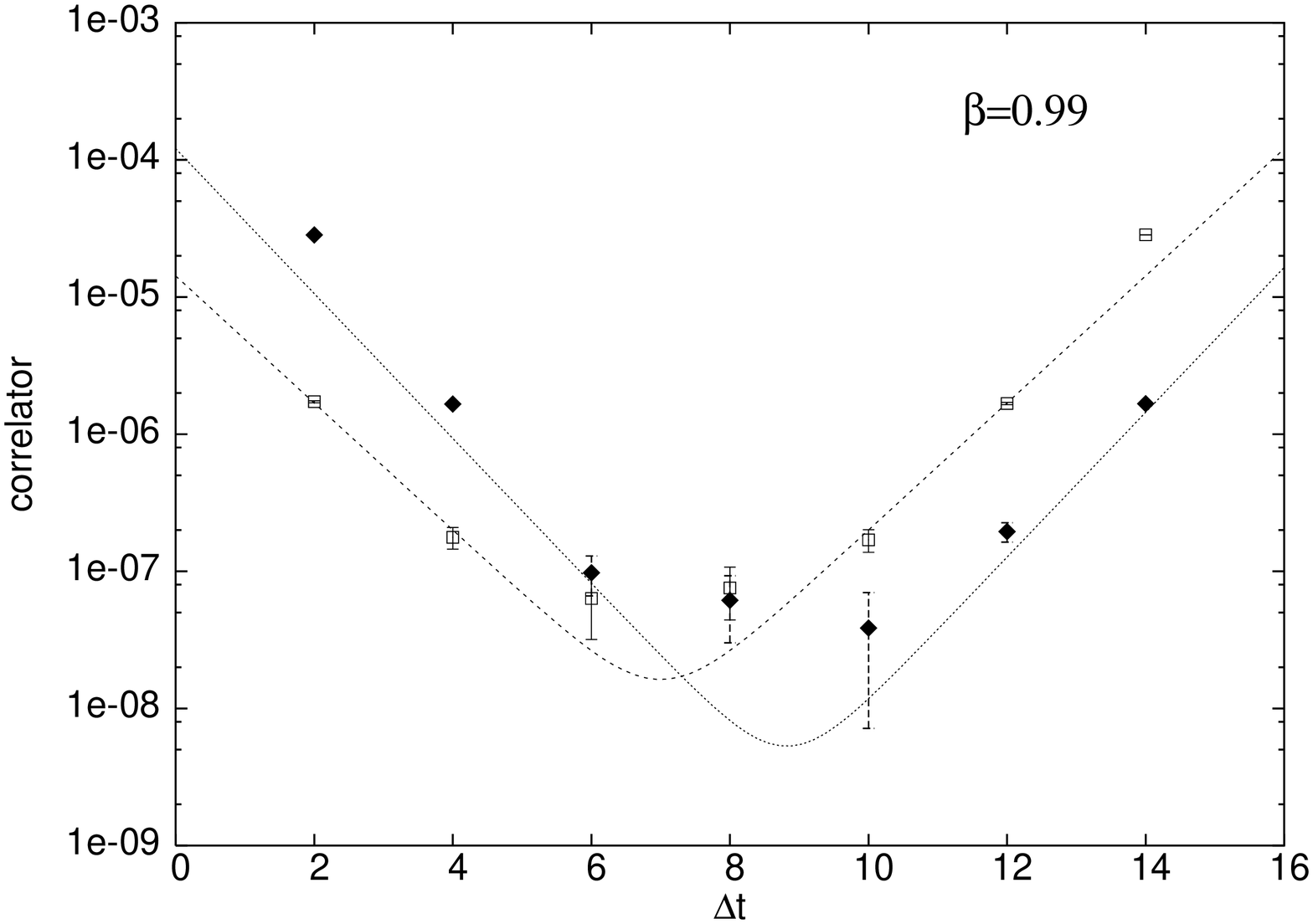,width=8truecm,angle=0}
\epsfig{file=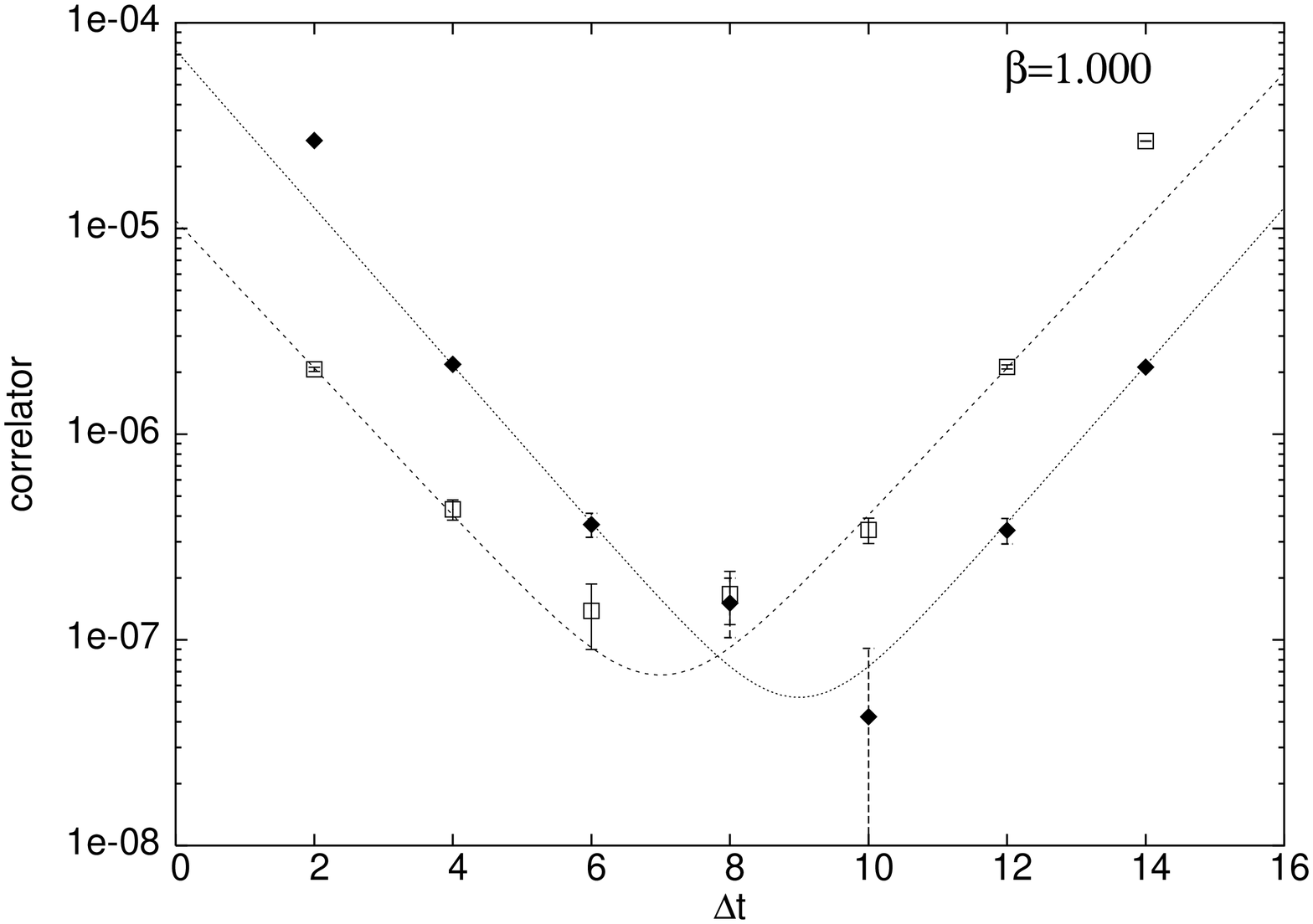,width=8truecm,angle=0}}
\mbox{\epsfig{file=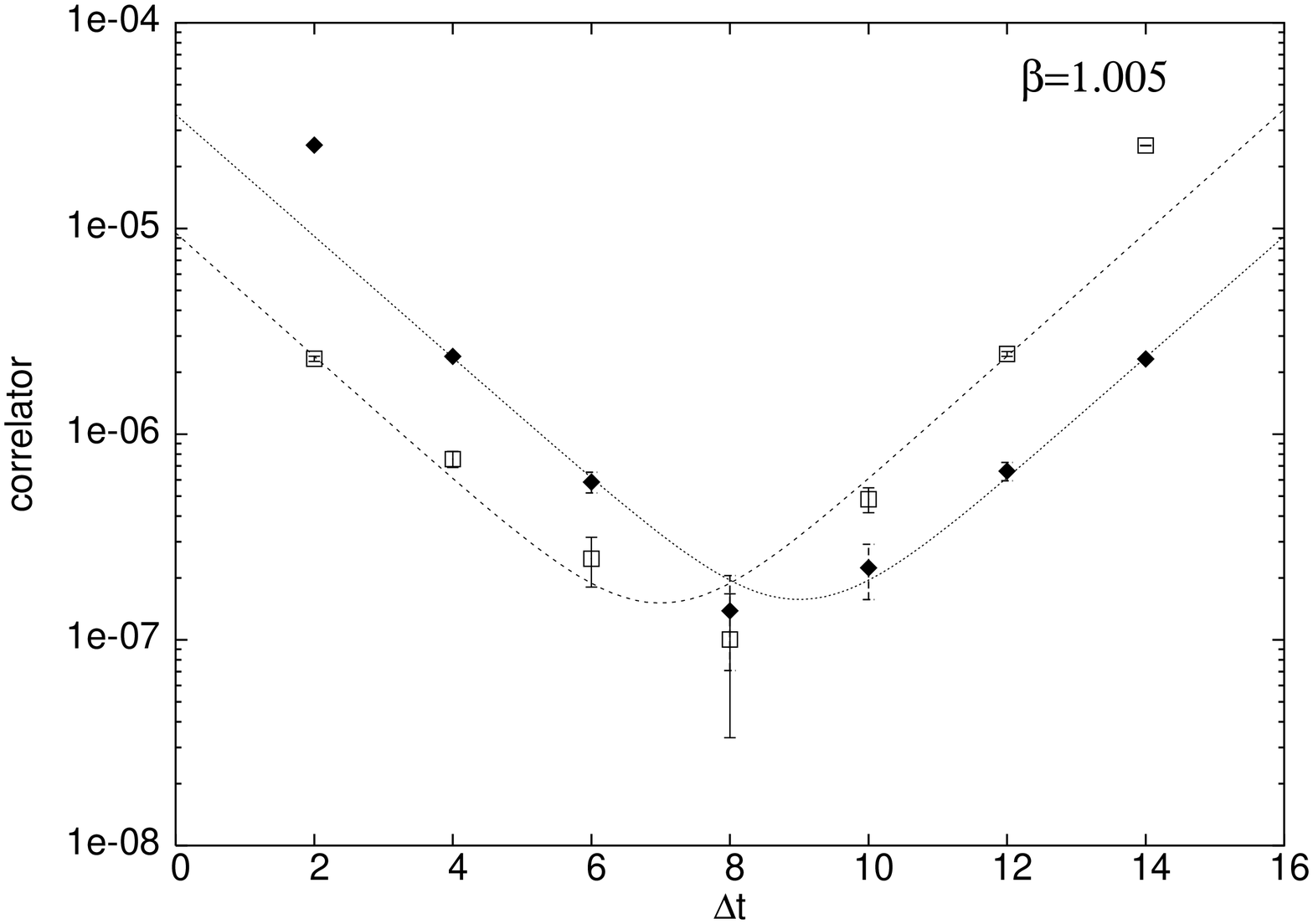,width=8truecm,angle=0}
\epsfig{file=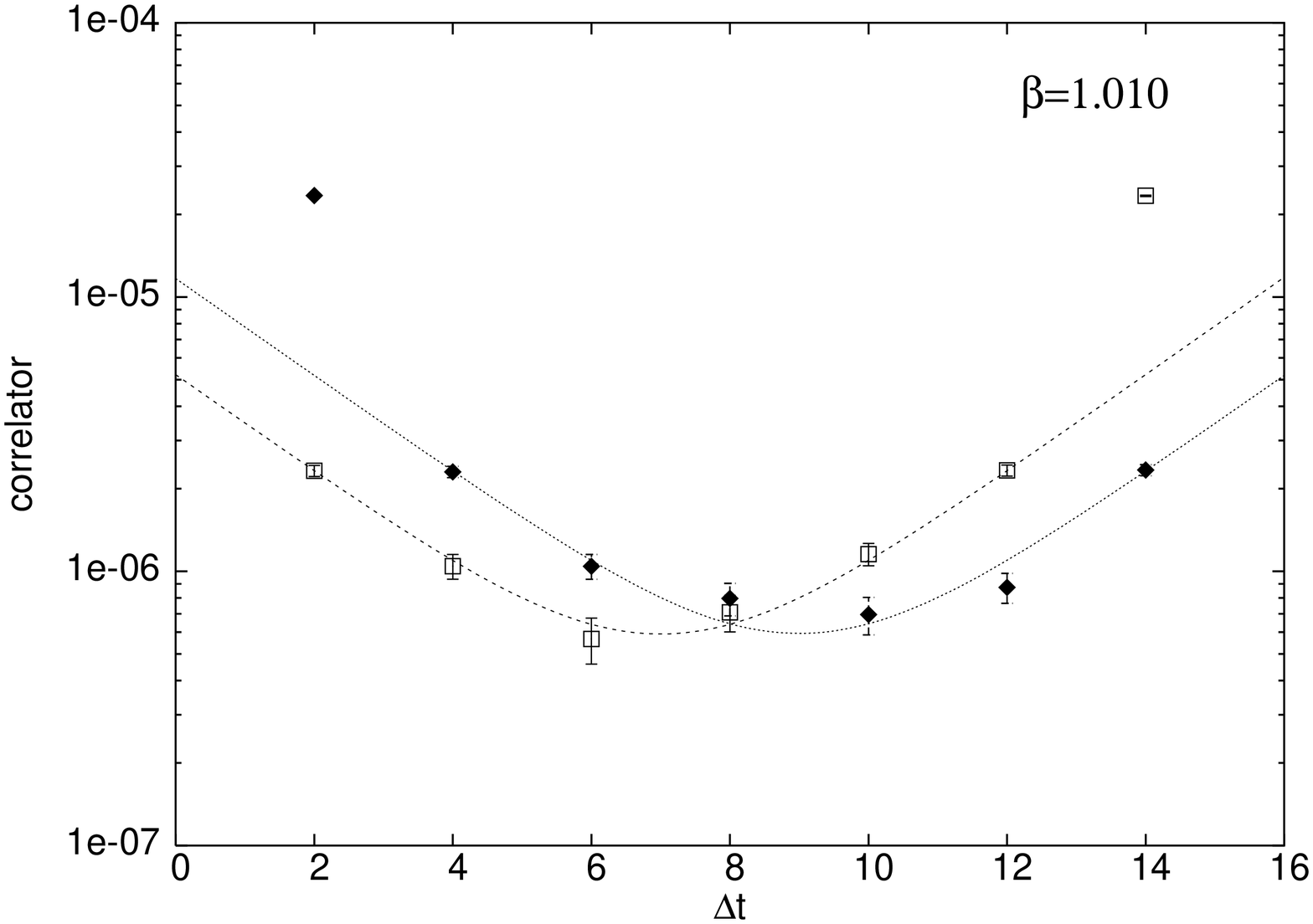,width=8truecm,angle=0}}
\caption{Scalar glueball correlators against $\Delta t$ for various values of $\beta$. 
Points denoted by $\square$ come from $\partial_t\partial_{t_0}^*
\langle C(t)C(t_0) \rangle$ while the ones denoted by $\blacklozenge$
come from $\partial_t^*\partial_{t_0} \langle C(t)C(t_0) \rangle $.}
\end{center}
\end{figure}

In these figures, points denoted by square ($\square$) come from the correlator
$\partial_t\partial_{t_0}^* 
\langle C(t)C(t_0) \rangle$ while the ones denoted by the filled diamond ($\blacklozenge$) 
come from the set 
$\partial_t^*\partial_{t_0} \langle C(t)C(t_0) \rangle $. At $\beta=1.000$
and $1.010$, where we could compare, our data is completely consistent with the
values reported in \cite{Stack,Oldphoton}.

Finally we plot the masses we have obtained from our simulations along with the prediction 
from the strong coupling expansion in fig. 3. 
The strong coupling expansion for the scalar glueball mass is
\cite{munster}
\be
m=-4\log u + \sum_{k=1}^{\infty}m_ku^k
\ee
where $u$ is given by $u=I_1(\beta)/I_0(\beta)$ for the group U(1)
and the coefficients $m_k$ till the eighth order are $m_2=\frac{3}{2}, m_4
=-\frac{793}{24}, m_6=-\frac{1783}{24}$ and $m_8=-\frac{1660309}{2880}$.
The coefficients of the odd powers of $k$ are zero upto this order.
From this figure we see that either the simulation points lie outside the
radius of convergence of the strong coupling expansion or the
higher order coefficients come with different signs.

\begin{figure}[htb]
\begin{center}
\mbox{\epsfig{file=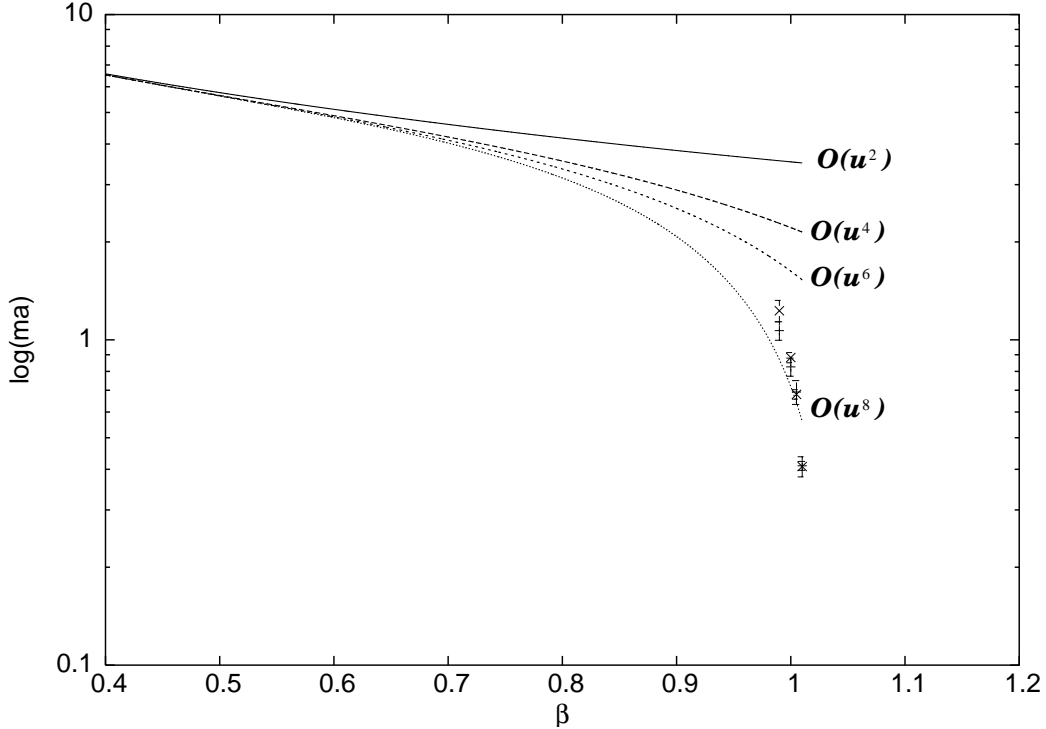,width=14truecm,angle=0}}
\caption{ Scalar glueball masses and the strong coupling expansions.}
\end{center}
\end{figure}

\subsection{Axial vector channel}

In the axial vector channel, we construct states with zero as well 
as with definite momenta. The explicit form of the correlator is given in
Eq.(\ref{vector}).
To obtain the effective masses from the correlators, we again fold the data 
about the symmetry point ($\Delta t=8$ in this case) and perform a correlated 
fit to
\be
f(\Delta t)=A \cosh(m_{\rm eff}\Delta t).
\ee
In table II we present the effective masses from the axial-vector
correlators along with the range of the fit, the momenta and the correlated
$\chi^2$ as obtained from the fits.
In fig. 4 we plot the zero momentum correlators at $\beta$ values 1.000 
($\blacktriangle$), 1.005 ($\blacksquare$) and 1.010 ($\blacktriangledown$).

\begin{figure}[htb]
\begin{center}
\mbox{\epsfig{file=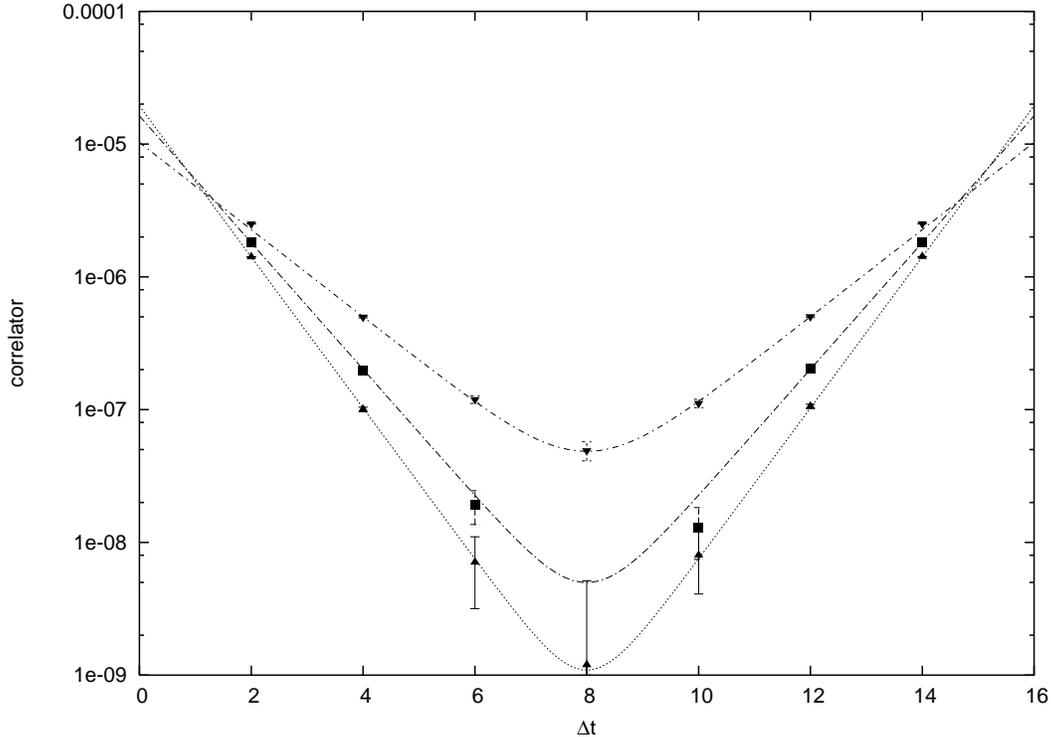,width=10truecm,angle=-90}}
\caption{Axial-vector correlators with zero momentum against $\Delta t$. 
$\beta$ =1.000 ($\blacktriangle$), 1.005 ($\blacksquare$) 
and 1.010 ($\blacktriangledown$)}
\end{center}
\end{figure}

\section{Discussion}

In section III we have seen that the momentum dependent axial vector correlator
is sensitive to the photon in the deconfined region. However to extract
a massless particle we need to subtract the momentum contribution to the 
effective mass and therefore specify the dispersion relation 
for this particle. As a first approximation we assume that the particle obeys 
the lattice free field dispersion relation \cite{Oldphoton}
\be\label{disp}
m_{\rm eff}^2=m_{\rm rest}^2+ \sum_{i=1}^3 (2-2\cos k_i), 
\ee
and check whether this assumption is consistent with our data or not.

\begin{figure}[htb]
\begin{center}
\mbox{\epsfig{file=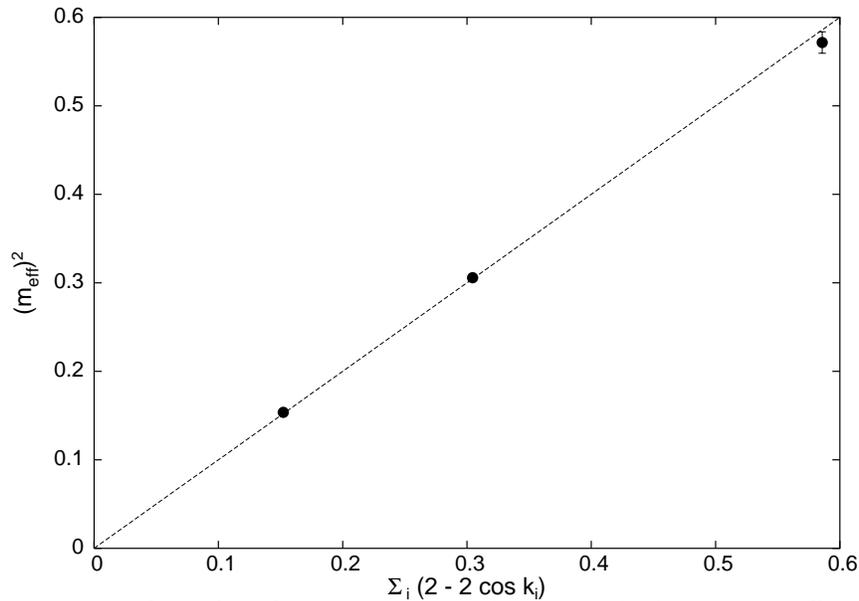,width=8truecm,angle=-90}}
\caption{Consistency of the free field dispersion relation Eq.(\ref{disp}) 
with the effective masses
from the axial-vector correlator at $\beta=1.015$. The momentum contribution to $m_{\rm eff}$
due to $(k_x,k_y,k_z)$ are plotted along the $x$-axis. }
\end{center}
\end{figure}

In fig. 5, the filled circle ($\bullet$) denotes the effective masses at $\beta=1.015$ for the 
momenta
$(k_x,k_y,k_z)=(2\pi/16,0,0), (2\pi/16,2\pi/16,0)$ and $(4\pi/16,0,0)$. The $x$-axis is
the contribution to $m_{\rm eff}$ due to these momenta, assuming the lattice free field 
dispersion
relation. The values
used in this plot are given in table II along with the effective masses. We have used 
values over the smaller range of $\Delta t$ (4 - 8) as that had a much lower $\chi^2/d.o.f$.
From the plot it is clear that within our statistical error bars, the 
data is completely consistent with a massless particle obeying the free field dispersion 
relation. While there is no justification to assume that this simplest relation holds close to
the phase transition point, our data shows that any deviation would have to be small. 

To see the behaviour of this massless particle at various values of $\beta$,
we have plotted the effective masses obtained from the momentum dependent
 axial-vector correlators
 in fig. 6. Here the triangles ($\vartriangle$) correspond to effective
masses obtained by
fitting over the whole range of $\Delta t$ while the inverted triangles ($\triangledown$) are the
masses obtained by  
dropping the smallest value of $\Delta t$. The filled circle ($\bullet$) has been obtained by
subtracting the
momentum contribution from the inverted triangles assuming free field dispersion relation and
the dotted lines are the 3 $\sigma $ bands for this set. The actual values along with the
1, 2 and 3 $\sigma $ limits are given in table III.

\begin{figure}[htb]
\begin{center}
\mbox{\epsfig{file=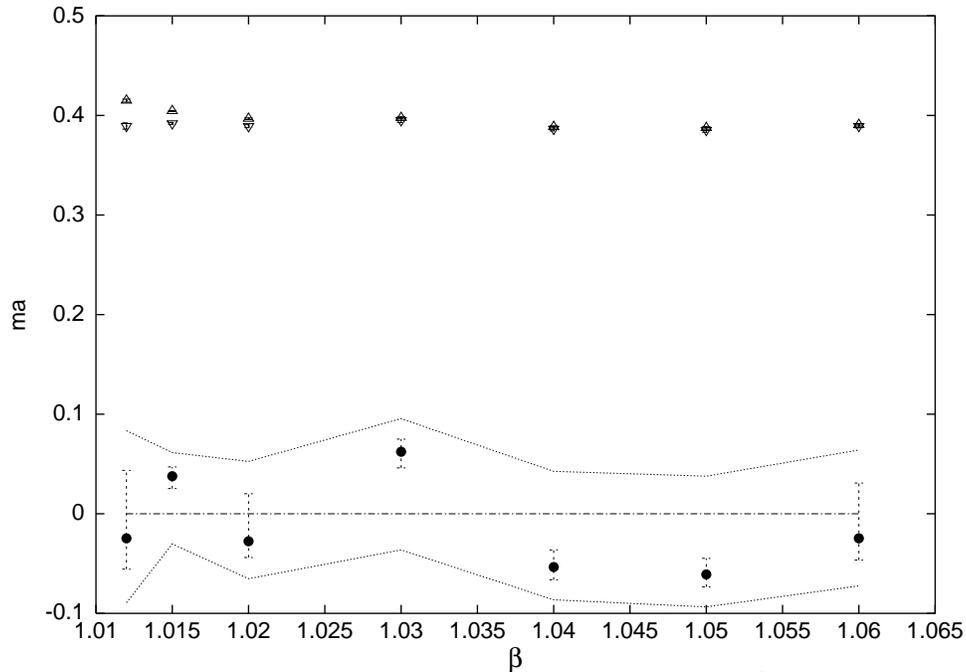,width=9truecm,angle=-90}}
\caption{Extracting the photon. $\vartriangle$ correspond to $m_{\rm eff}$ obtained by
fitting over the whole range of $\Delta t$ while $\triangledown$ are $m_{\rm eff}$ obtained by
dropping the smallest value of $\Delta t$. $\bullet$ has been obtained by
removing the momentum contribution assuming Eq.(\ref{disp}) and the dotted lines are the
3 $\sigma $ bands for this set. (Note that the zero-momentum correlator has no contribution
 from the photon. The dominant contribution comes from the correlator with the lowest possible 
non-zero momenta on the lattice.)} 
\end{center}
\end{figure}

From this figure it is clear that we are sensitive to the presence of higher states
up to $\beta=1.02$. Beyond this, dropping the smaller values of $\Delta t$ do not yield
significantly different values of the effective mass. Moreover within statistical 
fluctuations all the masses are consistent with zero.

\begin{figure}[htb]
\begin{center}
\mbox{\epsfig{file=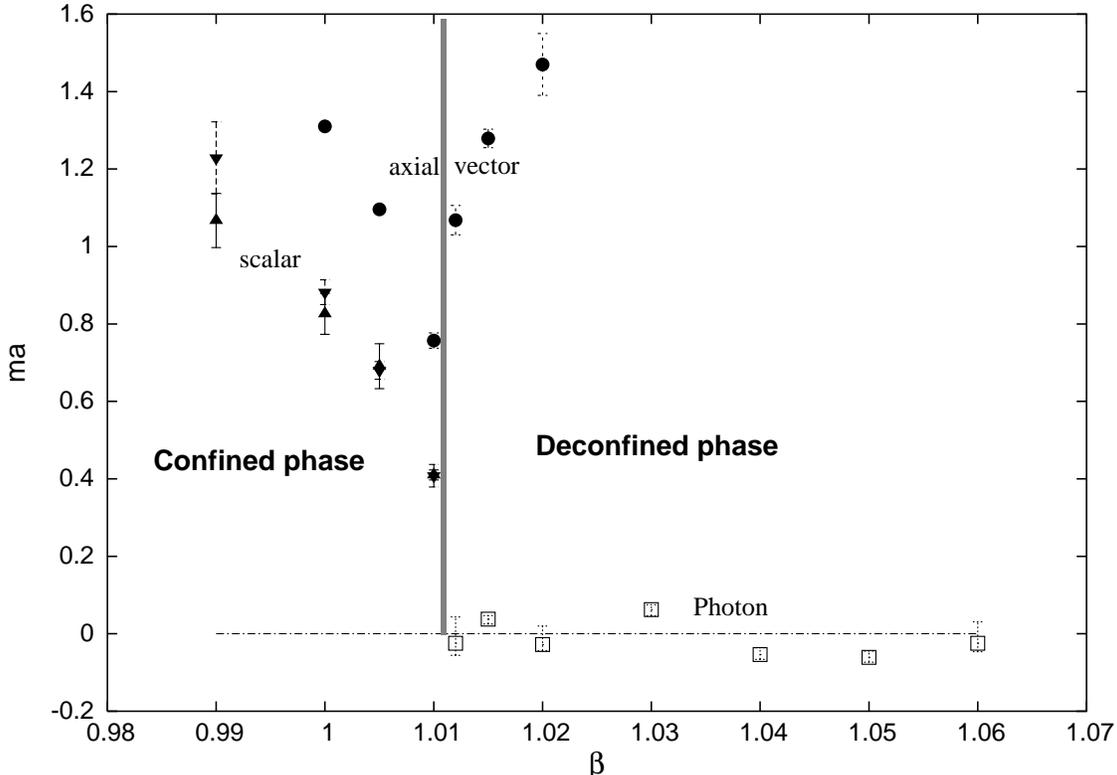,width=15truecm,angle=0}}
\caption{Consolidated plot of glueball masses.
$\blacktriangle$ and $\blacktriangledown$
are the scalar glueball masses from the sets $\partial\partial^* \langle CC
\rangle$ and $\partial^*\partial \langle CC \rangle $ respectively. The
$\bullet$ corresponds to the axial vector mass with zero momentum.
The $\square$ is the photon extracted from the axial vector correlator (See fig. 6).}
\end{center}
\end{figure}
Finally in fig. 7, we plot the masses from all the channels along with the region
where the phase transition is believed to take place.
The filled triangle ($\blacktriangle$) and the filled inverted 
triangle ($\blacktriangledown$)
are the scalar glueball masses from the sets $\partial\partial^* \langle CC
\rangle$ and $\partial^*\partial \langle CC \rangle $ respectively. The 
filled circle ($\bullet$)
corresponds to the axial vector mass where the momentum has been set to zero.
The square ($\square$) is the photon extracted from the axial vector correlator (fig. 6).
From this figure we see that on the confining 
side the scalar mass falls to about 0.41 in lattice units near the phase
transition while the axial vector mass at 0.76 lattice units is nearly twice that value.
However none of them seem to go to zero. Although we have not done a finite size 
analysis, our lattice volume itself is reasonably large : $(\sim 2 f\!m)^4$ near the 
phase transition. 
Also during equilibriation at $\beta=1.010$ we observed that the plaquette seemed to 
get stuck in a different phase for some time when we started from an ordered configuration.
Putting these two facts together 
we believe that our data points towards a first order phase 
transition which is in agreement with \cite{ALSN}.
In the deconfined region we see a remarkable difference between the zero
momentum and the non-zero momentum correlators. While the zero momentum correlators
vary over two orders of magnitude as the $\Delta t$ varies, the correlators with 
definite momenta hardly seem to change. In fact if we subtract the free field 
momentum dependence we get rest masses which are zero within errors. We believe 
this is numerical evidence for the photon in the deconfined phase.

The final picture that emerges from our study is that the general scenario presented in 
\cite{Oldphoton} does not change. The momentum dependent correlator indeed seems to 
indicate the presence of a zero mass particle in the deconfined phase and the axial
vector which comes from the zero momentum part seems to get heavy and decouple from the 
theory. We have carried out the 
analysis on a $16^4$ lattice and have obtained the masses from correlators which are separated
by 2 to 8 lattice spacings. This lets us disentangle the effect of the higher states 
to the effective masses of the lightest glueballs and therefore yields much more accurate 
results. In spite of such large temporal separations ($\sim 1f\!m$ in terms of our scale),
the multi-level scheme of measurements together 
with taking the derivatives to get rid of the non-zero vacuum expectation values, allow us 
to obtain accurate data. This demonstrates the power of this method which 
 should be applicable in the non-Abelian cases as well.

Our study also helps further consolidate the pattern of glueball masses in U(1) lattice gauge
theory and points towards a first order confinement deconfinement phase transition. Furthermore
we present strong evidence for a massless photon in the deconfined phase of the theory. 

\section{Acknowledgements}

The authors would like to express their gratitude to Peter Weisz 
and  Erhard Seiler for constant encouragement
and numerous discussions during the course of this work. We are also indebted to Martin
 L\"uscher for suggesting the use of derivatives and to Tom DeGrand for several useful comments
and help with the correlated fitting. Thanks are also due to Pierre Van Baal and Ferenc 
Niedermayer for a critical reading of the manuscript and several useful suggestions.

Finally PM would like to thank the Rechen Zentrum Garching and the 
Max-Planck-Institut for the computing facilities and YK and MK would like to thank SX5 at
RCNP, Osaka University where part of the computation was carried out.

\newpage

\begin{table}
\begin{center}
\begin{tabular}{cccccc}
&& \multicolumn{2}{c}{$\partial_t^*\partial_{t_0}\langle C(t)C(t_0)\rangle $} & 
\multicolumn{2}{c}{$\partial_t\partial_{t_0}^*\langle C(t)C(t_0)\rangle $} \\
$\beta$ & range of $\Delta t$& $m$ & $\chi^2/d.o.f$ & $m$ & $\chi^2/d.o.f$ \\
\hline 
0.990 & 2 - 8 & 1.414 (7) & 8.8/2 & 1.402 (7) & 21.51/2 \\
& 4 - 8 & 1.195 (55) & 2.8/1 & 1.085 (50) & 4.04/1 \\ \\
1.000 & 2 - 8 & 1.241 (8) & 38.01/2 & 1.2485 (80) & 60.73/2 \\
& 4 - 8 & 0.875 (41) & 0.53/1 & 0.812 (37) & 3.219/1 \\ \\
1.005 & 2 - 8 & 1.1545 (90) & 93.06/2 & 1.1455 (90) & 88.64/2 \\
& 4 - 8 & 0.682 (29) & 0.11/1 & 0.693 (29) & 0.127/1 \\ \\
1.010 & 2 - 8 & 1.075 (15) & 215.7/2 & 1.065 (14) & 227.8/2 \\
& 4 - 8 & 0.405 (22) & 4.26/1 & 0.410 (21) & 0.0039/1 
\end{tabular}
\caption{\label{scalarmass} Scalar glueball masses}
\end{center}
\end{table}

\begin{table}
\begin{center}
\begin{tabular}{ccccc}
$\beta$ & momenta & range of $\Delta t$& $m_{\rm eff}$ & $\chi^2/d.o.f.$ \\
\hline
1.000 & (0,0,0) & 2 - 8 & 1.31 (1) & 0.0013/2 \\ \\
1.005 & (0,0,0) & 2 - 6 & 1.096 (9) & 3/1 \\ \\
1.010 & (0,0,0) & 2 - 8 & 0.809 (5) & 5.31/2 \\
& & 4 - 8 & 0.757 (20) & 0.012/1 \\ \\
& (0,0,0) & 1 - 3 & 1.068 (38) & 2.26/1 \\
1.012 & ($2\pi/16$,0,0) & 2 - 8 & 0.4149 (16) & 80.0/2 \\
 & & 4 - 8 & 0.3894 (32) & 0.36/1 \\ \\
& (0,0,0) & 2 - 8 & 1.278 (24) & 5.7/2 \\
&& 2 - 6 & 1.279 (24) & 0.24/1 \\
& ($2\pi/16$,0,0) & 2 - 8 & 0.4044 (5) & 256.0/2 \\
1.015&& 4 - 8 & 0.3920 (10) & 3.43/1 \\ 
&($2\pi/16$,$2\pi/16$,0) & 2 - 8 & 0.5685 (10) & 31.84/2 \\
&& 4 - 8 & 0.553 (3) & 0.022/1 \\
& ($4\pi/16$,0,0) & 2 - 8 & 0.766 (2) & 1.59/2 \\
&& 4 - 8 & 0.756 (8) & 0.0069/1 \\ \\
& (0,0,0) & 2 - 8 & 1.47 (8) & 0.46/2 \\
1.020 & ($2\pi/16$,0,0) & 2 - 8 & 0.3968 (8) & 30.82/2 \\
&& 4 - 8 & 0.3892 (15) & 0.39/1 \\ \\
1.030 & ($2\pi/16$,0,0) & 2 - 8 & 0.3976 (11) & 3.06/2 \\
&& 4 - 8 & 0.3951 (22) & 1.4/1 \\ \\
1.040 & ($2\pi/16$,0,0) & 2 - 8 & 0.3885 (10) & 1.87/2 \\
&& 4 - 8 & 0.3865 (20) & 0.61/1 \\\\
1.050 & ($2\pi/16$,0,0) & 2 - 8 & 0.3875 (12) & 6.4/2 \\
&& 4 - 8 & 0.3854 (22) & 5.2/1 \\\\
1.060 & ($2\pi/16$,0,0) & 2 - 8 & 0.3905 (10) & 0.36/2 \\
&& 4 - 8 & 0.3894 (20) & 0.03/1

\end{tabular}
\caption{\label{vectormass} Mass from the axial-vector correlator.}
\end{center}
\end{table}

\begin{table}
\begin{center}
\begin{tabular}{ccr|rr|rr|rr}
$\beta$ & $m_{\rm eff}$ & $m_{\rm rest}$ & \multicolumn{2}{c|}{1 $\sigma$ limits} &
\multicolumn{2}{c|}{2 $\sigma$ limits} &\multicolumn{2}{c}{3 $\sigma$ limits} \\
\hline
1.012 & 0.3894 (32) & $-$0.0247 & $-$0.0556 & 0.0435 & $-$0.0745 & 0.0665 & $-$0.0894 & 0.0834 \\
1.015 & 0.3920 (10) & 0.0377 & 0.0253 & 0.0470 & $-$0.0119 & 0.0547 & $-$0.0303 & 0.0615 \\
1.020 & 0.3892 (15) & $-$0.0276 & $-$0.0439 & 0.0201 & $-$0.0556 & 0.0397 & $-$0.0652 & 0.0525 \\
1.030 & 0.3951 (22) & 0.0622 & 0.0461 & 0.0749 & 0.0201 & 0.0858 & $-$0.0362 & 0.0955 \\
1.040 & 0.3865 (20) & $-$0.0535 & $-$0.0663 & $-$0.0362 & $-$0.0770 & 0.0158 & $-$0.0864 & 0.0426
\\
1.050 & 0.3854 (22) & $-$0.0609 & $-$0.0735 & $-$0.0448 & $-$0.0841 & $-$0.0172 & $-$0.0935 &
0.0377 \\
1.060 & 0.3894 (20) & $-$0.0247 & $-$0.0465 & 0.0309 & $-$0.0609 & 0.0502 & $-$0.0724 & 0.0640
\end{tabular}
\caption{\label{photonmass} Level of significance of nonzero $m_{\rm rest}$ for the photon.}
\end{center}
\end{table}

\end{document}